\documentclass{aa}
\usepackage{psfig}
\begin{document}
\thesaurus{ 12
      (08.16.7;
       09.09.1) }
\title{On the velocity of the Vela pulsar}

\author{Vasilii Gvaramadze\inst{1,2,3}\thanks{{\it Address for
correspondence}: Krasin Str.,
19, ap. 81, Moscow, 123056, Russia; e-mail: vgvaram@mx.iki.rssi.ru}}

\institute{
Abastumani Astrophysical Observatory, Georgian Academy of Sciences, A.Kazbegi ave.
2-a, Tbilisi, 380060, Georgia
\and Sternberg State Astronomical Institute, Universitetskij Prospect 13,
Moscow, 119899, Russia
\and Abdus Salam International Centre for Theoretical Physics, Strada Costiera 11,
P.O. Box 586, 34100 Trieste, Italy
}

\date{Received / accepted }

\maketitle

\begin{abstract}
It is shown that if the shell of the Vela supernova
remnant is responsible for nearly all the scattering of the Vela pulsar,
then the scintillation and proper motion velocities of the pulsar can only be
reconciled with each other in the case of nonzero transverse velocity of
the scattering material. A possible origin of large-scale
transverse motions in the shell of the Vela supernova remnant is discussed.


\end{abstract}

\section{Introduction}
%
The \object{Vela pulsar} is one of the best studied radio pulsars and was the first
one found  to be associated with the supernova remnant (SNR) (\cite{la}).
In spite of its relative proximity to the Earth, there is still no
consensus on the value of its (transverse) velocity, which is connected
with the yet unsolved problem of the distance to the Vela pulsar/Vela SNR. The
first attempt to estimate the pulsar velocity was made by \cite{bi},
whose optical measurments gave an upper limit on the pulsar proper
motion ($< 60 \, {\rm mas}\,{\rm yr}^{-1}$). However, even the maximum admissible
value of the
pulsar proper motion was found to be too low to explain the pulsar offset from the
apparent geometrical centre of the Vela SNR, which questions the
pulsar/SNR association. Later, it was recognized (e.g. \cite{se}, \cite{a})
that the real extent of the Vela SNR is much larger than
was accepted in early studies, so now there can be no doubt
that the Vela pulsar and the Vela SNR are the remnants of the same supernova
explosion. However,  this association has caused  some problems in estimating
the pulsar velocity.

\cite{ws} questioned for the first time the
'canonical' value of the distance to the Vela SNR of 500 pc given by Milne
(\cite{m1}; see also \cite{tml}) and suggested that this distance should be
reduced to some smaller value ($\simeq 250$ pc). Since that time
many additional arguments in support of this suggestion have been put forward
(\"Ogelman et al. \cite{ok}, Oberlack et al. \cite{o}, \cite{j1},
\cite{a}, \cite{boc}, Cha et al. \cite{ch2}, \cite{ch1})\footnote{We
critically analysed these
arguments (\cite{g3}) and came to the conclusion that there are
no weighty reasons to revise the 'canonical' distance of 500 pc. }.
One of the arguments (proposed by
\"Ogelman et al. (\cite{ok}) and repeated by Oberlack et al. (\cite{o})
and Cha et al. (\cite{ch2})) was
based on the comparison of the Vela pulsar
transverse velocity (inferred
from the new estimate of the pulsar proper motion
by \"Ogelman et al. (\cite{ok})) with the scintillation
velocity (reported by Cordes (\cite{c1})).
It is known that proper motion velocities of pulsars show significant
correlation with pulsar velocities inferred from interstellar scintillation
measurements (e.g. \cite{ly}, Gupta \cite{gu1}). Therefore, assuming that the
scintillation velocity $V_{\rm iss}$ is a 'true' value of the pulsar transverse
velocity, \"Ogelman et al. (\cite{ok}) suggested that $V_{\rm iss} = 53\pm 5 \, {\rm
km}\,{\rm s}^{-1}$ (Cordes \cite{c1}) could be reconciled with the proper motion
$\mu \simeq 38 \pm 8 \, {\rm mas}\,{\rm yr}^{-1}$ if the distance
to the Vela pulsar (and the Vela SNR) is about $290 \pm 80$ pc.
The distance
reduction might be even more dramatic if one takes the recent high-precision
estimate
of the Vela pulsar proper motion ($52 \pm 3 \, {\rm mas}\,{\rm yr}^{-1}$)
obtained by De Luca et al. (\cite{d}; see also \cite{n}).
The situation with the distance to the Vela SNR was 'improved' after
Gupta et al. (\cite{gu2})
showed that the scintillation
velocity calculation formula used by Cordes (\cite{c1})  underestimates
$V_{\rm iss}$ by factor of 3.
The revised value of the scintillation
velocity of the Vela pulsar of $152 \, {\rm km}\,{\rm s}^{-1}$
better corresponds to the proper motion velocity of
$123 \, {\rm km}\,{\rm s}^{-1}$ (for the distance
to the pulsar of 500 pc and $\mu = 52 \, {\rm mas}\,{\rm yr}^{-1}$).
However, the real situation is a bit more complicated.

It should be noted that the calculations of Gupta (\cite{gu1})
were based on the assumptions that the scattering material
is concentrated in
a thin screen and that the screen is placed
midway between the observer and the
pulsar. Although the first assumption is realistic, the second one is not
suitable in the case of the Vela pulsar. Indeed, it is believed that the
scattering irregularities responsible for the enhanced scattering of the Vela pulsar
(\cite{b}) are localized in a thin screen rather than uniformly distributed along
our line of sight to the pulsar (\cite{b}, \cite{le}; see
also \cite{w}), and that the
scattering screen resides close to the pulsar and could be associated with the
shell of the Vela SNR (Desai et al. \cite{de}, \cite{tc}, \cite{gw1}; see
however Gwinn et al. \cite{gw2},\cite{gw3} and
cf. \cite{c2}). The asymmetrical location of the
screen implies (see Sect. 2) that the actual value of the scintillation
velocity should be considerably larger than that given by Gupta
(\cite{gu1}) and that the
scintillation velocity is not equal to the proper motion velocity.

In this
paper, we show that if the scattering of the Vela pulsar indeed occurs in the
shell of the Vela SNR then the scintillation velocity could be
reconciled with the pulsar proper motion velocity only if the scatterer has a
nonzero transverse velocity. A possible origin of large-scale transverse motions in
the Vela SNR's shell is discussed.

\section{Scintillation and proper motion velocities}

The scintillation velocity for an asymmetrically placed thin scattering screen is
(Gupta et al. \cite{gu2}, Gupta \cite{gu1}):
\begin{equation}
\label{1}
V_{\rm iss} = 3.85\times 10^4 \,{(\nu _{{\rm d},{\rm MHz}} D_{\rm kpc} x)^{1/2}
\over f_{\rm GHz}
t_{\rm d}} \, {\rm km}\,{\rm s}^{-1} \, ,
\end{equation}
where $\nu _{{\rm d},{\rm MHz}}$ and $t_{\rm d}$ are the scintillation bandwidth and
the time-scale measured respectively in MHz and seconds, $D_{\rm kpc}$ is the
distance
from observer to pulsar in kpc, $x=D_{\rm o} /D_{\rm p}$, $D_{\rm o}$ and $D_{\rm
p}$ are
the distances from observer to screen and from screen to pulsar, $f_{\rm GHz}$ is
the frequency of observation in units of GHz.  For $\nu_{{\rm d},{\rm MHz}} =
0.001$, $t_{\rm d} = 5.6$, $f_{\rm GHz} =1$ (Cordes \cite{c1}), $D_{\rm kpc} = 0.5$, and
assuming that
$x=1$, one finds for the Vela pulsar that $V_{\rm iss} = 152 \, {\rm km}\,{\rm
s}^{-1}$
(Gupta \cite{gu1}). As we mentioned in Sect. 1, Desai et al. (\cite{de}) showed that the
scattering screen is close to the pulsar. Assuming that $D=500$ pc, they
found that $D_{\rm o} /D \simeq 0.81$, and that this value could be increased up to
0.96 if $5 \%$ of the scattering of the Vela pulsar is due to the effect of the
\object{Gum Nebula}. The latter value of $D_{\rm o} /D$ is expected if the
scattering material is mainly concentrated in the shell of the Vela SNR of
angular diameter of $5^{\degr}$ (the figure
accepted in early studies of the Vela SNR).
Assuming that the Vela SNR's shell is
indeed the main scatterer of the Vela pulsar
and using the currently adopted angular
size of the Vela SNR of $\simeq 7^{\degr}$, one has $D_{\rm o} /D \simeq 0.94$ or
$x=15.7$, and correspondingly $V_{\rm iss} \simeq 600 \, {\rm
km}\,{\rm s}^{-1}$. In the observer's reference frame, the scintillation velocity is
connected with the pulsar proper motion velocity
\begin{equation}
\label{2}
V_{\rm pm} = 4.74 \mu D_{\rm kpc} \, {\rm km}\,{\rm s}^{-1} \, ,
\end{equation}
where $\mu$ is measured in mas\,${\rm yr}^{-1}$, by the following
relationship (cf. Gupta et al. \cite{gu2}, \cite{c2}):
\begin{eqnarray}
\label{3}
V_{\rm iss} &=& \Big[x^2 V_{\rm pm} ^2 -2x(1+x)V_{\rm pm} V_{{\rm scr},\parallel}
 \nonumber \\
      &+&(1+x)^2 V_{{\rm scr},\parallel} + (1+x)^2 V_{{\rm scr},\perp} ^2 \Big]^{1/2} \, ,
\end{eqnarray}
where $V_{{\rm scr},\parallel}$  and $V_{{\rm scr},\perp}$ are the
components of the
transverse
velocity of the screen, correspondingly, parallel and perpendicular to the vector
of the pulsar proper motion velocity. In (\ref{3}) we neglected small contributions from
the differential Galactic rotation and the
Earth's orbital motion around the Sun. If $V_{\rm scr} =0$,
one has $V_{\rm pm} =
V_{\rm iss} /x \simeq 38 \, {\rm km}\,{\rm s}^{-1} (D_{\rm kpc} =0.5)$,
i.e. about 3 times smaller than
that from eq. (\ref{2}). These velocity estimates could be
reconciled only if the distance to the Vela pulsar is $D_{\rm kpc} =
0.05(x/15.7)^{-1}$, which is too small to be likely. The pulsar, however,
could be placed at its 'canonical' distance if $V_{\rm scr} \neq 0$.

\begin{figure*}
\centerline{\psfig{figure=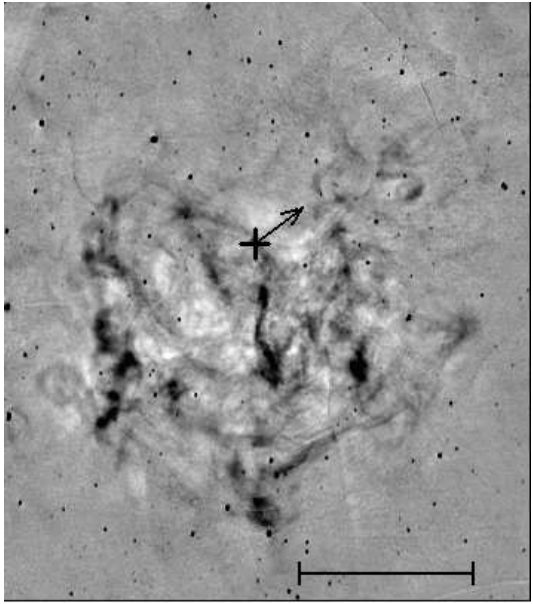,width=7cm}}
\caption[]{The 843 MHz image of the central part of the Vela SNR (adopted from \cite{bo}).
Position of the Vela pulsar is indicated by a cross. The
arrow shows the direction of the pulsar proper motion velocity (\cite{ba}).
North is up, east at left. The horizontal bar is $1^{\degr}$ long.}
\label{f1}
\end{figure*}

Fig.~\ref{f1} shows the 843 MHz image (\cite{bo}) of the central part of
the Vela SNR, known as the
radio source \object{Vela X} (Milne \cite{m1}).
A considerable fraction of the radio
emission
from Vela\,X originates in filamentary structures, one of which crosses the Vela
pulsar position.
This filament (or rather its
part to the south of the pulsar) is known as a radio counterpart of the Vela X-ray
'jet' discovered by \cite{mo}. In Gvaramadze (\cite{g1}, \cite{g2}), we
found that some of radio filaments of the Vela\,X and the X-ray 'jet' show a fairly
good correlation with optical filaments, and concluded that the 'jet' is a dense
filament in the Vela SNR's shell, which is projected by chance near the line of
sight to the Vela pulsar. We also suggested that filamentary structures
visible throughout the Vela SNR in radio, optical and X-ray ranges have a common
nature\footnote{This suggestion implies that the radio source Vela\,X is a part of
the Vela SNR's shell (see also \cite{mm}, \cite{m2}; but see e.g.
\cite{f}, \cite{bo} and \cite{ch3} for a
different point of view).} and that
their origin is connected with projection effects in the Rayleigh-Taylor unstable
shell of the remnant. The Rayleigh-Taylor instability results from the impact of the
supernova ejecta/shock with the pre-existing wind-driven shell created by the
supernova progenitor star (see Gvaramadze \cite{g2}), and induces in the shell
large-scale transverse motions (laterally expanding domelike deformations of the
shell). The existence of laterally expanding deformations naturally explains
(Gvaramadze \cite{g2}) the
'unusual' velocity field inferred by \cite{j3} from the study of
absorption lines in spectra of background stars (see also \cite{j2},
\cite{da}). The high-velocity absorption features were found not only in the
central part of the Vela SNR but also near the edges of the remnant, which suggests
that the expansion velocity of the shell deformations has comparable radial and
transverse components. The same conclusion follows from the interpretation of UV
spectra of face-on and edge-on shock waves in the Vela SNR (\cite{ra}). A characteristic
expansion velocity of shell deformations inferred from
the absorption data and UV spectra is about $100 \, {\rm km}\,{\rm s}^{-1}$, while
some portions of the shell expand with much higher velocities.

Proceeding from the above, we suggest that the radio filament projected
on the Vela pulsar is a large-scale deformation of the Vela SNR's shell viewed
edge-on, and
that this deformation has a significant transverse velocity. We assume that the
deformation lies on the approaching side of the Vela SNR's shell and suggest
that the turbulent material associated with the shell deformation is responsible
for the scattering of the Vela pulsar (cf. Desai et al. \cite{de}). The line of
sight extent of the scattering material could be estimated to be $\simeq 1.2 -
1.7$ pc (for $D = 500$ pc) given that the width of the filament is $\simeq
2^{'} - 3^{'}$ (\cite{m2}, \cite{bo}) and assuming that the
characteristic size of the shell deformation is $\simeq 40^{'} - 50^{'}$.
The geometry (the curvature) of the filament projected on the Vela pulsar
suggests that this part of the shell expands in the northwest direction,
i.e. just parallel to the vector of the pulsar proper motion velocity
(\"Ogelman et al. \cite{ok}, \cite{ba}). For $V_{{\rm scr},\perp} =0$,
one has from (\ref{3}) that
\begin{equation}
\label{4}
V_{\rm scr} = V_{{\rm scr},\parallel}
\simeq {xV_{\rm pm} \pm V_{\rm iss} \over x+1} \, .
\end{equation}
For the transverse velocity of the pulsar of $\simeq 120 \, {\rm
km}\,{\rm s}^{-1}$ (i.e. for $D = 500$ pc) and for $x=15.7$ and $V_{\rm iss} \simeq
600
\, {\rm km}\,{\rm s}^{-1}$, one has that $V_{\rm scr}$ is 80 or
$150 \,{\rm km}\,{\rm s}^{-1}$. The first estimate is quite reasonable,
while the second,  though not impossible, looks less likely. It should be
noted, however, that eq. (\ref{4}) gives reasonable values for $V_{\rm scr}$ not only for
$D=500$ pc. E.g. for $D=250$ pc, one finds that $V_{\rm scr}$ is equal
to 30 or $80 \, {\rm
km}\,{\rm s}^{-1}$. From this, we conclude that
the scintillation data taken alone do not allow us to put meaningful
limits on the distance to the Vela pulsar and therefore to get a
reliable estimate of the transverse velocity of the pulsar. The forthcoming direct
measurement of the distance to the Vela pulsar through its parallactic displacement
(see De Luca et al. \cite{d}) will solve the problem.

\section{Discussion}

The main assumption made in this paper is that the enhanced scattering of
the Vela pulsar is due to the effect of the Vela SNR. This assumption is
based on the result of Desai et al. (\cite{de}) that the scattering material
could be associated with the shell of the Vela SNR (see Sect. 2). In this
case, the parameter $x$ has a fixed value, which depends only on the
angular size of the Vela SNR's shell, and therefore is independent of the
distance to the pulsar.

It should be noted, however, that the result of Desai et al.
(\cite{de}) was derived from use of scintillation observables (see
\cite{bl}, \cite{gw1}). The referee (J.Cordes) pointed out that
the quite large uncertainties in observables result in uncertainty
in the value of the parameter $x$. He attracted our attention to
the papers by Gwinn et al. (\cite{gw2},\cite{gw3}), from which
follows that the value of $x$ could be much smaller than that
adopted in our paper. The values of $x$ given in these papers
(respectively, $x=2.7 (D/0.5 \, {\rm kpc})^{-1}$ and $x=1.5 (D/0.5
\, {\rm kpc})^{-1} )$ imply that the scattering material cannot be
connected with the Vela SNR. We now discuss this possibility.

For $x=2.7 (D/0.5 \, {\rm kpc})^{-1}$ (Gwinn et al. \cite{gw2}), one
obtains\footnote{For the sake of simplicity,
we use here the same values of $\nu _{\rm d} , t_{\rm d}$ and $f$ as in
Sect. 2.} $V_{\rm iss} =250 \, {\rm km}\,{\rm s}^{-1}$. Note that
now $V_{\rm iss}$ is independent of $D$ since
$x\propto D^{-1}$ (see e.g. \cite{gw1} and eq. (\ref{1})). One can see
that $V_{\rm iss}$ could be reconciled with $V_{\rm pm}$ only if $V_{\rm
scr} \neq 0$: $V_{\rm scr} = 157$ or $-22 \, {\rm km}\,{\rm s}^{-1}$ for
$D=0.5$ kpc; $V_{\rm scr} = 91$ or $-13 \, {\rm km}\,{\rm s}^{-1}$ for
$D=0.25$ kpc . These transverse velocities are too high to be attributed
to the expansion of the shell of the Gum Nebula\footnote{The enhanced
scattering of the Vela pulsar was originally ascribed to the Gum Nebula
(\cite{b}; see also \cite{c3}).} (cf. \cite{re} with \cite{wsj}),
though it is not impossible that they could
characterize the expansion of a foreground small-scale H\,II region
projected by chance on the Vela pulsar.

For $x=1.5 (D/0.5 \, {\rm kpc})^{-1}$ (Gwinn et al. \cite{gw3}),
one obtains $V_{\rm
iss} = 186 \, {\rm km}\,{\rm s}^{-1}$. One can show that just for this
value of $x$ (or more exactly for $x=(152/123)^2 =1.53$) the transverse
velocity of the screen could be equal to zero (see eq. (\ref{4})),
and therefore $V_{\rm pm} =V_{\rm iss} /x$.

Although these estimates show that the situation with the scattering screen is
indeed quite uncertain, we believe that the enhanced scattering of the
Vela pulsar is most likely connected with the shell of the Vela SNR. An
argument in support of this belief is the fact that the elongated
scattering disk of the Vela pulsar (Gwinn et al. \cite{gw2}) is nearly
perpendicular to the magnetic field of the radio filament projected on the
Vela pulsar.

\begin{acknowledgements}
I am grateful to D.C.-J.Bock for providing the electronic version of
the image of the central part of the Vela SNR, and to J.Cordes (the
referee) for useful suggestions and comments.
\end{acknowledgements}

\end{document}